\documentclass[floatfix,twocolumn,showpacs,amsmath,amssymb]{revtex4}
\usepackage{graphicx,amsmath,amssymb}
\usepackage{dcolumn}
\usepackage{bm}
\usepackage{psfrag}

\newcommand{\nn}{\nonumber\\}
\newcommand{\gvec}[1]{\vec{#1}}

\begin{document}
\title{Resonant forcing of nonlinear systems of differential equations}
\author{Vadas Gintautas}
\altaffiliation[Also at ]{T-7 and the Center for Nonlinear Studies, Theoretical Division, Los Alamos National Laboratory, Los Alamos NM 87545, USA}
\email{vgintau2@uiuc.edu}
\author{Alfred W. H\"{u}bler}
\email{a-hubler@uiuc.edu}
\affiliation{Center for Complex Systems Research, Department of Physics, University of Illinois at Urbana-Champaign, Urbana, Illinois 61801, USA}
\date{\today}
\begin{abstract}
We study resonances of nonlinear systems of differential equations, including but not limited to the equations of motion of a particle moving in a potential.  We use the calculus of variations to determine the minimal additive forcing function that induces a desired terminal response, such as an energy in the case of a physical system.  We include the additional constraint that only select degrees of freedom be forced, corresponding to a very general class of problems in which not all of the degrees of freedom in an experimental system are accessible to forcing.  We find that certain Lagrange multipliers take on a fundamental physical role as the effective forcing experienced by the degrees of freedom which are not forced directly.  Furthermore, we find that the product of the displacement of nearby trajectories and the effective total forcing function is a conserved quantity.  We demonstrate the efficacy of this methodology with several examples.
\end{abstract}
\pacs{05.45.Xt, 05.45.-a}

\maketitle
{\bf  
Resonance in nonlinear systems is an important topic that has been explored in depth~\cite{wargitsch95b,foster07,gintautas08}.  Resonance in a linear system is defined to be a maximum response amplitude when driven by a signal at a specific frequency.  In this case something about the forcing function, namely, the forcing frequency, mirrors something about the system, its natural frequency.  Previous work on driven damped nonlinear oscillators demonstrated that such a system will achieve a maximum amplitude when the forcing dynamics matches the time-reversed dynamics of the same system without forcing~\cite{hubler92}.  Again there is a relationship between the natural dynamics of the system and the dynamics of the drive.  In this paper we derive the most efficient resonant forcing function possible for a very general class of systems, namely, systems which can be described by coupled first-order differential equations, including systems which exhibit chaos~\cite{lorenz63}.  In the methodology we present, there is no restriction on the degrees of freedom which may be forced, so it is possible to compute the resonant forcing of, say, one of two coupled oscillators.   In such a system, only one of the four degrees of freedom would be forced; this was not possible previously.  We show that optimal forcing functions may be used for system parameter identification via resonance spectroscopy.  Furthermore, conservation laws in closed systems usually correspond to a fundamental symmetry.  In this paper we show that an open dissipative system subject to optimal resonant forcing has a special conserved quantity and a corresponding symmetry.  This conserved quantity is the dot product of the separation of nearby trajectories and the effective forcing experience by all degrees of freedom.  
}
\section{Introduction}
\label{sec:contin_intro}

There has been extensive work on sinusoidally driven nonlinear oscillators in the contexts of synchronization~\cite{eisenhammer90}, stochastic resonance~\cite{kapral93,bulsara05} and nonlinear response phenomena~\cite{morton70,siddiqi05}.  Resonance phenomena of nonlinear systems due to aperiodic and chaotic forcing functions~\cite{mallick05,foster07} has been less studied, but results from work in this area indicate that generally a nonlinear oscillator will have a greater response when driven with the correct aperiodic signal rather than a sinusoidal one.  A related topic is system identification via resonance curves of nonlinear systems~\cite{krempl92} and periodically driven chaotic systems~\cite{ruelle86}.  Plapp and H\"{u}bler~\cite{plapp90} and others~\cite{wargitsch95b} have used the calculus of variations to show that a special class of aperiodic driving forces can achieve a large energy transfer to a nonlinear oscillator.  Such nonsinusoidal resonant forcing functions yield a high signal-to-noise ratio which can be used for high-resolution system identification~\cite{chang91}.  In a recent paper, Gintautas, Foster, and H\"{u}bler~\cite{gintautas08} explored resonant forcing of time-discrete chaotic dynamics.  In this work, we extend their method to time-continuous systems of ordinary differential equations, including but not limited to, the equations of motion of a particle in a potential, and show that the optimal forcing function induces a desired response more efficiently than a sinusoidal forcing function.

Systems of first order differential equations are ubiquitous in modern science and engineering.  Furthermore, any higher order differential equation, such as an equation of motion for a Hamiltonian system,  or system of equations may be cast as a set of first order equations.  Systems of differential equations have been used to model a rich variety of systems, ranging from complex networks~\cite{li07} to jet flow~\cite{uleysky07}, to give two very recent examples.  In these cases the correct model accurately reproduces the natural dynamics of the system.  In other cases, it is important not only to correctly model the unperturbed dynamics of the system but also to be able to control or influence these dynamics.  

In this paper, we present a methodology for determining the resonant forcing of a system of first order differential equations in which only select degrees of freedom are forced.  This is motivated by the difficulty or impossibility of forcing all of the degrees of freedom in certain experiments.  For example, consider a physical oscillator in which it is possible to directly force the position $x$ but not the velocity $\dot{x}$.  Therefore, the method we present may be applied to a very general class of problems.  We show analytically that the resonant forcing functions are closely related to the unperturbed dynamics of the system in that the product of the displacement of nearby trajectories and the effective total forcing function is a conserved quantity.  We also show that the optimal forcing for a damped oscillator moving in a potential is proportional to the time reflected dynamics of the corresponding unperturbed system; this is the ``principle of the dynamical key'' explored by H\"{u}bler et al.~\cite{hubler92,wargitsch95b}.  Furthermore, we find that certain Lagrange multipliers take on a fundamental physical role as the efficiency of the forcing function and the effective forcing experienced by the degrees of freedom which are not forced directly.  We demonstrate the efficacy of the methodology with several examples. Since the method we present is general and requires only access to one degree of freedom, nearly any system that is accurately modeled using a system of first order equations can also in principle be controlled efficiently, including systems which exhibit chaos~\cite{lorenz63}.  

\section{General formulation}
\label{sec:contin_optforce}
We begin with a multidimensional first order system with forcing:
\begin{equation}
\dot{\vec{x}}=\vec{f}\bigl(\vec{x}\bigr)+\vec{F},
\label{eq:contin_xdyn}
\end{equation}
where $\vec{x}=\vec{x}(t)\in\mathbb{R}^{d}$ denotes the state of the $d$-dimensional system at time $t$, and $\vec{F}=\vec{F}(t)\in\mathbb{R}^{d}$ denotes the forcing function at time $t$.  This system has $d$ degrees of freedom.  We seek to minimize the total forcing effort, that is, the integral of the magnitude of $\vec{F}$ from $t=0$ to $t=\tau$, which we define to be the constant $\bar{F}^{2}$:
\begin{equation}
    \bar{F}^{2} \equiv \frac{1}{2}\int^{\tau}_{0}\bigl[\vec{F}(t)\cdot\vec{F}(t)\bigr]dt.
\label{eq:contin_Fmagdefn}
\end{equation}
Here the terminal time $\tau$ is a free parameter.  We require that $0\leq d_{u}<d$ degrees of freedom be unforced.  Without loss of generality, we choose to order the variables so that $x_{1},\ldots,x_{d_{u}}$ are unforced and $x_{d_{u}+1},\ldots,x_{d}$ are forced.  Thus we will require that
\begin{alignat}{2}
F_{i}(t)=0,&\qquad&\text{for $i=1,\ldots,d_{u}$ and $0\leq t\leq\tau$},
\label{eq:contin_fzero}
\end{alignat}
where $F_{i}(t)$ is the $i$th component of $\vec{F}(t)$.  This problem can be solved by a variation of the functional $S=\int_{0}^{\tau}L_{g}dt$.  The Lagrange function $L_{g}$ is given by
\begin{equation}
L_{g}=L(\vec{x},\dot{\vec{x}},\vec{F},t)+ \lambda K(\vec{x},\dot{\vec{x}},t)\delta_{D}(t-\tau),
\label{eq:contin_lagrangian}
\end{equation}
where $\delta_{D}(t-\tau)$ is the Dirac delta function and $\lambda$ is a constant Lagrange multiplier.  The function $K$ is a generalized boundary condition for $t=\tau$ and represents a constraint at the terminal time.  We will require that $K$ be in the form
\begin{equation}
    K\bigl[\vec{x}(t),\dot{\vec{x}}(t),t\bigr]=0 \qquad\text{at $t=\tau$}.
    \label{eq:contin_Kcondition}
\end{equation}
In light of the above constraints, $L$ is given by
\begin{equation}
    L=\frac{1}{2} \vec{F}\cdot\vec{F} + \vec{F}\cdot\gvec{\Gamma} + \gvec{\mu}\bigl(t\bigr)\cdot\bigl[\dot{\vec{x}}-\vec{f}\bigl(\vec{x}\bigr)-\vec{F}\bigr].
\label{eq:contin_Ldefn}
\end{equation}
Because the equation of motion in Eq.~\eqref{eq:contin_xdyn} is a nonintegral constraint, $\gvec{\mu}(t)$ is a time dependent Lagrange multiplier.  We have defined the vector $\gvec{\Gamma}(t)\equiv\sum^{d_{u}}_{j=1}\gamma_{j}(t)\hat{\mathbf{e}}_{j}$, where $\gamma_{1}(t),\ldots,\gamma_{d_{u}}(t)$ are time dependent Lagrange multipliers and $\hat{\mathbf{e}}_{j}$ is the unit basis vector in the direction of $x_{j}$.  This term represents the constraint that certain degrees of freedom not be forced.  Thus the Lagrange problem is 
\begin{equation}
\delta S=\delta\int_{0}^{\tau}L + \lambda K\delta_{D}(t-\tau)dt,
\label{eq:contin_varproblem}
\end{equation}
where $\delta S$ is the variation of $S$.  Following Wargitsch and H\"{u}bler~\cite{wargitsch95b}, we derive the Euler-Lagrange equations for this problem (in which the terminal time is a free parameter) in the Appendix.  The equations of motion are 
\begin{gather}
\frac{\partial L}{\partial x_{i}}-\frac{d}{dt}\Bigl(\frac{\partial L}{\partial \dot{x}_{i}}\Bigr)=0,\label{eq:contin_EOMeq1}\\
\frac{\partial L}{\partial F_{i}}=0,\label{eq:contin_EOMeq2}
\end{gather}
for $i=1,\ldots,d$.   At the upper boundary, for $t=\tau$,
\begin{align}
&\frac{\partial K}{\partial \dot{x}_{i}}=0,\label{eq:contin_UBeq1}\\
\lambda &\frac{\partial K}{\partial x_{i}}+\frac{\partial L}{\partial \dot{x}_{i}}=0,\label{eq:contin_UBeq2}\\
\lambda &\frac{\partial K}{\partial t}+L-\sum_{i=1}^{d}\Bigl(\dot{x}_{i}\frac{\partial L}{\partial \dot{x}_{i}}\Bigr)=0,\label{eq:contin_UBeq3}
\end{align}
for $i=1,\ldots,d$.  At the lower boundary, we have the initial condition $\vec{x}(0)$.  The equations of motion yield:
\begin{gather}
\mathbf{J}^{T}\gvec{\mu} + \dot{\gvec{\mu}} = 0, \label{eq:contin_firstFeqn}\\
\vec{F} + \gvec{\Gamma} - \gvec{\mu} = 0, \label{eq:contin_secondFeqn}
\end{gather}
where $J_{ij}=\bigl(\partial f_{i}/\partial x_{j}\bigr)\bigr|_{\vec{x}\left(t\right)}$ is the Jacobi matrix evaluated at $\vec{x}(t)$.  The superscript $T$ indicates the transpose operator.  We now define the quantity
\begin{equation}
\vec{G}\equiv\vec{F}+\gvec{\Gamma},
\label{eq:contin_Gdefn}
\end{equation}
noting that the components of $\vec{G}$ in the direction of unforced degrees of freedom are the corresponding components of $\gvec{\Gamma}$ and the components of $\vec{G}$ in the direction of forced degrees of freedom are equal to the corresponding components of $\vec{F}$.  When we solve for the Lagrange multiplier $\gvec{\mu}$ using Eq.~\eqref{eq:contin_secondFeqn},
\begin{equation}
\gvec{\mu}(t) = \vec{G}(t),
\end{equation}
then Eqs.~\eqref{eq:contin_EOMeq1} and \eqref{eq:contin_EOMeq2} reduce to simply
\begin{equation}
\dot{\vec{G}} = -\mathbf{J}^{T}\vec{G}.
\label{eq:contin_GandJ}
\end{equation}
From Eq.~\eqref{eq:contin_Gdefn} and \eqref{eq:contin_GandJ} we identify $\vec{G}$ as the effective total forcing function; it reduces to the optimal forcing $\vec{F}$ when we remove the constraint in Eq.~\eqref{eq:contin_fzero}.  We identify the Lagrange multipliers $\gamma_{1},\ldots\gamma_{d_{u}}$ to be the effective forcing experienced by the degrees of freedom $j$ for which $F_{j}=0$; this changes the trajectories of these degrees of freedom via the coupling in $\vec{f}(\vec{x})$ rather than direct additive forcing via $\vec{F}$.  We now further simplify the upper boundary conditions in Eqs.~\eqref{eq:contin_UBeq2} and \eqref{eq:contin_UBeq3}, which become
\begin{align}
    \lambda \frac{\partial K}{\partial x_{i}} &= -G_{i}(\tau), \label{eq:contin_UBeq2_G}\\
    \lambda \frac{\partial K}{\partial t} & = \frac{1}{2}\vec{F}(\tau)\cdot\vec{F}(\tau) + \vec{G}(\tau)\cdot\vec{f}\bigl[\vec{x}(\tau)\bigr], \label{eq:contin_UBeq3_G}
\end{align}
for $i=1,\ldots,d$.  These boundary conditions along with the initial condition $\vec{x}(0)$ and Eqs.~\ref{eq:contin_xdyn} and \ref{eq:contin_GandJ} form a complete boundary value problem which, in principle, may be solved analytically or numerically to determine $\gvec{\Gamma}(t)$ and $\vec{F}(t)$ for $0\leq t\leq\tau$.

The control is stable if, on average, the displacement of nearby trajectories decreases.  Consider a trajectory given by Eq.~\eqref{eq:contin_xdyn}, and a nearby trajectory given by $\dot{\vec{x}}^{\prime}=\vec{f}\bigl(\vec{x}^{\prime}\bigr)+\vec{F}$, where $\vec{x}$ and $\vec{x}^{\prime}$ are related by $\gvec{\epsilon}\equiv\vec{x}-\vec{x}^{\prime}$.  If we Taylor expand $\vec{f}\bigl(\vec{x}\bigr)$ for small $\gvec{\epsilon}$, we obtain
\begin{equation}
\dot{\gvec{\epsilon}}=\mathbf{J}\gvec{\epsilon}.\label{eq:contin_conserveddyn}
\end{equation}
Multiplying both sides of the transpose of Eq.~\eqref{eq:contin_GandJ} by $\gvec{\epsilon}$, we have $\dot{G}^{T}\epsilon=-G^{T}\mathbf{J}\epsilon$.  Using Eq.~\eqref{eq:contin_conserveddyn}, this becomes $\dot{\vec{G}}\cdot\gvec{\epsilon}=-\vec{G}\cdot\dot{\gvec{\epsilon}}$, or
\begin{equation}
\frac{d}{dt}\bigl(\gvec{\epsilon}\cdot\vec{G}\bigr)=0,
\end{equation}
a quantity that is invariant for all $t$.  We define this to be the conserved quantity $P$:
\begin{equation}
P\equiv\gvec{\epsilon}\cdot\vec{G},
\label{eq:contin_conserved}
\end{equation}
and note that $P$ depends on the observables $\vec{x}$ and $\vec{F}$ as well as the Lagrange multipliers in $\gvec{\Gamma}$, which we have identified as the effective indirect forcing of certain degrees of freedom.  This further reinforces the idea that $\vec{G}$ represents the effective forcing experienced by the system, taking into account the coupling via $\vec{f}(\vec{x})$.  Note that $P$ is conserved even if the unperturbed dynamics is chaotic or periodic.  We can independently show that $P$ is a conserved quantity using the invariance of the Lagrangian.  Consider the transformation,
\begin{align}
&\vec{x}\rightarrow\vec{x}+\gvec{\epsilon},\\
&\dot{\vec{x}}\rightarrow\dot{\vec{x}}+\dot{\gvec{\epsilon}}.
\end{align}
Under this transformation, the variation of the Lagrangian is in the form $L\rightarrow L+\delta L$, with
\begin{equation}
    \delta L = \gvec{\mu}\cdot\bigl(\dot{\gvec{\epsilon}}-\mathbf{J}\gvec{\epsilon}\bigr).
\end{equation}
Using Eq.~\eqref{eq:contin_conserveddyn}, $\delta L=0$ and we immediately see that the Lagrangian is invariant under this transformation.  Noether's theorem~\cite{marinho07} states that if $\delta L$ for a given transformation can be written as a total derivative of a function $U$, that is, $\delta L =dU/dt$, then there is a corresponding conserved quantity $j$ (called Noether's current):
\begin{equation}
    j=\gvec{\epsilon}\cdot\frac{\partial L}{\partial\dot{\vec{x}}}-U.
\end{equation}
In this case, $\delta L = dU/dt = 0$, so $U$ is some constant $c$, and it follows that
\begin{equation}
    \frac{d}{dt}j = \frac{d}{dt}\left(\gvec{\epsilon}\cdot\frac{\partial L}{\partial \dot{\vec{x}}}-c\right)=0.
\end{equation}
From Eqs.~\eqref{eq:contin_Ldefn} and \eqref{eq:contin_secondFeqn}, we have $\partial L/\partial \dot{\vec{x}}=\gvec{\mu}= \vec{G}$ and recover $dP/dt=0$.  Conservation laws in closed systems usually correspond to a fundamental symmetry.  Here we have shown that an open dissipative system subject to optimal resonant forcing has a special conserved quantity and a corresponding symmetry.

\section{Examples}
\subsection{One-dimensional damped oscillator}
We now illustrate the methodology with several examples.  Consider first a one-dimensional damped driven oscillator,
\begin{equation}
\ddot{x}+\eta\dot{x}+\frac{\partial V}{\partial x}=F(t),
\label{eq:contin_oscillator}
\end{equation}
where $\eta$ is the coefficient of linear damping and $V(x)$ is a time-independent potential.  The energy of the oscillator, $E(t)=\dot{x}(t)^{2}/2+V[x(t)]$, will provide a constraint at $t=\tau$.  We first consider a system of two coupled first-order equations:
\begin{equation}
\begin{pmatrix}
\dot{x}_{1}\\
\dot{x}_{2}
\end{pmatrix}=\begin{pmatrix}
x_{2}\\
-\eta x_{2}-\tfrac{\partial V}{\partial x_{1}}
\end{pmatrix}+\begin{pmatrix}
F_{1}\\
F_{2}
\end{pmatrix},
\label{eq:contin_matrixoscillator}
\end{equation}
with initial conditions
\begin{align}
    &x_{1}(0) = x_{0},\label{eq:contin_x1initcondOsc}\\
    &x_{2}(0) = v_{0}, \label{eq:contin_x2initcondOsc} 
\end{align}
and require that only $F_{2}$ be forced, that is, $F_{1}(t)=0$ for all $t$.  We will first solve for the forcing function in terms of $\vec{x}(t)$, then explore the meaning of the conserved quantity for this type of system.

\subsubsection{Equations of motion and general solution} 
Accordingly, $\gvec{\Gamma}(t)=\gamma(t)\hat{\mathbf{e}}_{1}$, where we have defined $\gamma(t)\equiv\gamma_{1}(t)$.  This system is equivalent to Eq.~\eqref{eq:contin_oscillator} if we identify $F_{2}(t)=F(t)$.  The Jacobi matrix for this system is
\begin{equation}
\mathbf{J}=\begin{pmatrix}
0&1\\
-\tfrac{\partial^{2}V}{\partial x_{1}^{2}}&-\eta
\end{pmatrix}.
\end{equation}

For the upper boundary condition, we will use
\begin{equation}
    K\bigl[\vec{x}(t),\dot{\vec{x}}(t),\tau\bigr] \equiv \frac{1}{2}x^{2}_{2}(t)+V\bigl[x_{1}(t)\bigr] - E = 0,\label{eq:contin_Kforosc}
\end{equation}
where $E$ is a constant energy value we wish the oscillator to attain at $t=\tau$.  Eq.~\eqref{eq:contin_GandJ} gives equations of motion for $F_{2}(t)$ and $\gamma(t)$:
\begin{align}
&\dot{\gamma}(t) = F_{2}(t) \frac{\partial^{2}V}{\partial x_{1}^{2}}, \label{eq:contin_FEOMOsc}\\
&\dot{F}_{2}(t) = \eta F_{2}(t) - \gamma(t). \label{eq:contin_gammaEOMOsc}
\end{align}
At the upper boundary, evaluating Eqs.~\eqref{eq:contin_UBeq2_G} and \eqref{eq:contin_UBeq3_G} gives
\begin{align}
    &\gamma(\tau) = -\lambda\frac{\partial V}{\partial x_{1}}\Bigr|_{t=\tau},\label{eq:contin_UBeval2Oscx1}\\
    &F_{2}(\tau) = -\lambda x_{2}(\tau),\label{eq:contin_UBeval2Oscx2}\\
    &\frac{1}{2} F^{2}_{2}(\tau) + \gamma x_{2}(\tau) = F_{2}(\tau) \biggl[\frac{\partial V}{\partial x_{1}}\biggr|_{t=\tau} + \eta x_{2}(\tau) \biggr]. \label{eq:contin_UBeval3Osc}
\end{align}
Eqs.~\eqref{eq:contin_UBeval2Oscx1}, \eqref{eq:contin_UBeval2Oscx2}, and \eqref{eq:contin_UBeval3Osc} can be solved for $\lambda$, $\gamma(\tau)$, and $F_{2}(\tau)$ in terms of $x_{1}(\tau)$ and $x_{2}(\tau)$ to provide explicit upper boundary conditions:
\begin{align}
    \lambda &= -2\eta,\\
    F_{2}(\tau) &= 2 \eta x_{2}(\tau), \label{eq:contin_FexplicitBCOsc}\\
    \gamma(\tau) &=\eta \frac{\partial V}{\partial x_{1}}\biggr|_{t=\tau}, \label{eq:contin_gammaexplicitBCOsc}
\end{align}
Eqs.~\eqref{eq:contin_matrixoscillator}--\eqref{eq:contin_x2initcondOsc}, \eqref{eq:contin_FEOMOsc}, \eqref{eq:contin_gammaEOMOsc}, \eqref{eq:contin_FexplicitBCOsc}, and \eqref{eq:contin_gammaexplicitBCOsc} form a well-posed boundary value problem.  Eliminating $\gamma$ from Eqs.~\eqref{eq:contin_FEOMOsc} and \eqref{eq:contin_gammaEOMOsc} gives an equation of motion for $F_{2}$: 
\begin{equation}
\ddot{F}_{2}(t)-\eta\dot{F}_{2}(t)+F_{2}(t) \frac{\partial^{2}V}{\partial x_{1}^{2}}=0.
\end{equation}
A trial solution for Eqs.~\eqref{eq:contin_oscillator} and \eqref{eq:contin_FEOMOsc} is given in the form
\begin{equation}
F_{2}(t)=\alpha\eta x_{2}(t).\label{eq:contin_Fx2sol}
\end{equation}
Using Eq.~\eqref{eq:contin_FexplicitBCOsc} , we find that this is a valid solution only for $\alpha=2$.  This is the same result as calculated by Wargitsch and H\"{u}bler~\cite{wargitsch95b} using a different formulation.

\subsubsection{Conserved quantity for one-dimensional damped oscillator}
Using this solution for $\vec{F}(t)$, we consider the conserved quantity $P$.  Using Eq.~\eqref{eq:contin_conserveddyn}, we obtain the following equation of motion for $\epsilon$:
\begin{equation}
\ddot{\epsilon}(t)+\eta\dot{\epsilon}(t)+\epsilon(t)\frac{\partial^{2}V}{\partial x_{1}^{2}}=0,
\label{eq:contin_epsEOMOsc}
\end{equation}
where we have defined $\epsilon\equiv\epsilon_{1}$ and eliminated $\epsilon_{2}=\dot{\epsilon}_{1}$.  This equation of motion is valid on the domain $0\leq t\leq \tau$.  We substitute Eq.~\eqref{eq:contin_Fx2sol} into Eq.~\eqref{eq:contin_oscillator} and operate on the resulting equation with an additional time derivative (henceforth for this example we will use $x_{1}\rightarrow x$ and $x_{2}\rightarrow \dot{x}$):
\begin{equation}
    \dddot{x}(t)-\eta\ddot{x}(t)+\dot{x}(t)\frac{\partial^{2}V}{\partial x^{2}}\biggr|_{t=t}=0.
\label{eq:contin_dxEOMOsc}
\end{equation}
Under the transformation $t\rightarrow \tau-t$ (from which it follows that $d/dt \rightarrow -d/dt$), this equation becomes
\begin{equation}
    \dddot{x}(\tau-t)+\eta\ddot{x}(\tau-t)+\dot{x}(\tau-t) \frac{\partial^{2}V}{\partial x^{2}}\biggr|_{t=\tau-t}=0,
\label{eq:contin_minusdxEOMOsc}
\end{equation}
where the second partial derivative $\partial^{2}V/\partial x^{2}$ is evaluated at $t=\tau-t$.  This equation is precisely in the same form as Eq.~\eqref{eq:contin_epsEOMOsc}, the equation of motion for $\epsilon$.  Furthermore, it is valid on the same domain, namely, $0\leq t\leq\tau$.  Therefore we identify
\begin{equation}
    \epsilon(t) = A \dot{x}(\tau-t) \propto F_{2}(\tau-t),
\label{eq:contin_epssol}
\end{equation}
with $A$ an arbitrary constant.  We use Eqs.~\eqref{eq:contin_GandJ}, \eqref{eq:contin_conserved}, \eqref{eq:contin_oscillator}, and \eqref{eq:contin_Fx2sol} to write $P$ in terms of $x$:
\begin{equation}
    P = \eta \dot{x}(t)\dot{x}(\tau-t)  + \dot{x}(t) \frac{\partial V}{\partial x}\biggr|_{t=\tau-t}
   -\dot{x}(\tau-t) \frac{\partial V}{\partial x}\biggr|_{t=t} ,\label{eq:contin_Psym}
\end{equation}
which is completely symmetric under the transformation $t\rightarrow \tau-t$.  We have absorbed any multiplicative constants into $P$.  Since $P$ does not change in time, Eq.~\eqref{eq:contin_Psym} must hold for $t=\tau/2$:  
\begin{equation}
    P=\eta \dot{x}\bigl(\tfrac{\tau}{2}\bigr)^{2}=\frac{d}{dt}\Bigl[\frac{1}{2}\dot{x}^{2}+V(x)\Bigr]\biggr|_{t=\frac{\tau}{2}} = \frac{dE(t)}{dt}\biggr|_{t=\frac{\tau}{2}}.
\end{equation}
Here $dE(t)/dt$ is the instantaneous rate of energy energy change of the oscillator and is equal to the instantaneous rate of energy transfer of the force $F_{2}(t)$.  When evaluated at $t=\tau/2$, it is equal to the conserved quantity $P$.
\subsection{Explicit example: isotonic harmonic oscillator}
\begin{figure}[htb]
\centering
 \psfrag{ome}{$\omega$}
 \psfrag{Ratio of forcing effort}{\hspace{10pt}$\bar{F}^{2}_{\text{optimal}}/\bar{F}^{2}_{\text{sinusoidal}}$}
 \psfrag{1.98}{1.98}
 \psfrag{1.99}{1.99}
 \psfrag{2.00}{2.00}
 \psfrag{2.01}{2.01}
 \psfrag{2.02}{2.02}
 \psfrag{2.03}{2.03}
 \psfrag{2.04}{2.04}
 \psfrag{0.24}{0.24}
 \psfrag{0.28}{0.28}
 \psfrag{0.32}{0.32}
 \psfrag{0.36}{0.36}
 \psfrag{0.40}{0.40}
    \includegraphics[width=0.45\textwidth]{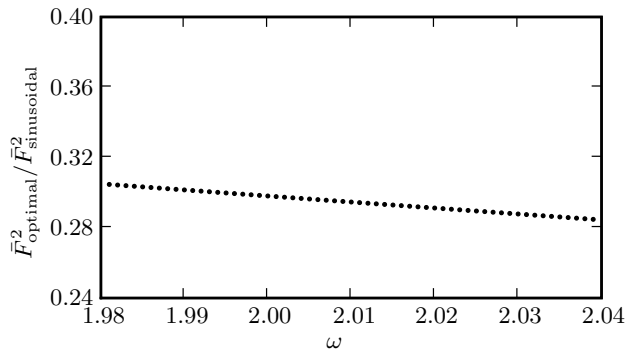}
    \caption{Ratio of total forcing effort for optimal forcing to that of sinusoidal forcing.  Since the optimal forcing is more efficient, this ratio is always less than $1$ (dashed line).  Here, $x(0) = 1.0$, $\dot{x}(0)=0.01$, $\eta = 0.3$, and $E=2.0$.  A different value of $\tau$ was calculate for each value of $\omega$ plotted. \label{fig:contin_compare}}
\end{figure}

We now illustrate the above formulation with several explicit examples.  Consider a forced isotonic harmonic oscillator of the form of Eq.~\eqref{eq:contin_oscillator}, with a potential $V=x^{2} \omega^{2}/2 +k/x^{2}$.  Thus the equation of motion of the oscillator is 
\begin{equation}
    \ddot{x}+\eta\dot{x}+\omega^{2}x-\frac{2k}{x^{3}}=F(t).
\label{eq:contin_isotonic}
\end{equation}
We will use generalized initial conditions, with $x(0)=x_{0}$ and $\dot{x}(0)=v_{0}$.  This potential represents a harmonic oscillator with a centripetal barrier~\cite{carinena05,chalykh05} and the corresponding equation of motion for $\eta = 0$ and no forcing is a particular case of the Pinney-Ermakov equation~\cite{pinney50}.  The unforced $\eta = 0$ case is is an example of a nonlinear isochronous system, that is, the amplitude of the oscillations of the solution are independent of the frequency.  For these examples, however, we will consider the system with damping so that $\eta$ is left as a free parameter.  First, we examine the case where $k=0$, corresponding to a simple damped harmonic oscillator.  For optimal forcing, $F(t)=2\eta \dot{x}(t)$, and the solution for $x(t)$ is
\begin{align}
    x(t)&=e^{\eta t/2}\biggl[x_{0}\cosh{\tfrac{t}{2}\sqrt{\eta^{2}-4 \omega^{2}}}\nn
    &+\frac{(2 v_{0}-\eta x_{0})}{\sqrt{\eta^{2}-4 \omega^{2}}} \sinh{\tfrac{t}{2}\sqrt{\eta^{2}-4 \omega^{2}}}\biggr].
\end{align}
It follows that
\begin{align}
    F(t)&=2 \eta\dot{x}(t)=2\eta e^{\eta t/2}\biggl[v_{0} \cosh{\tfrac{t}{2}\sqrt{\eta^{2}-4 \omega^{2}}}\nn
    &\qquad+\frac{(\eta v_{0}-2\omega^{2} x_{0})}{\sqrt{\eta^{2}-4 \omega^{2}}} \sinh{\tfrac{t}{2}\sqrt{\eta^{2}-4 \omega^{2}}}\biggr].\label{eq:contin_harmonicF2}
\end{align}
The solution for $\epsilon(t)$ is
\begin{align}
    \epsilon(t)&=\epsilon(0)e^{-\eta t/2}\cosh{\tfrac{t}{2}\sqrt{\eta^{2}-4 \omega^{2}}}\nn
    &+\bigl[2 \dot{\epsilon}(0)+\eta \epsilon(0)\bigr]\sinh{\tfrac{t}{2}\sqrt{\eta^{2}-4 \omega^{2}}},
\end{align}
where the initial conditions for $\epsilon$ at $t=0$ are $\epsilon(0)$ and $\dot{\epsilon}(0)$.  The conserved quantity can be computed exactly to be
\begin{equation}
    P=\omega^{2}x(0)\epsilon(0)+\dot{x}(0)\dot{\epsilon}(0).
    \label{eq:contin_explicitP}
\end{equation}
We can compare the effectiveness of this forcing function to that of sinusoidal driving.  For a given $\omega$, we compute the time $\tau$ such that the energy of the oscillator under optimal forcing [see Eq.~\eqref{eq:contin_harmonicF2}] is equal to the desired value given in Eq.~\eqref{eq:contin_Kforosc}.  Then we force the same oscillator using instead $F_{\text{sinusoidal}}(t) = A \sin \omega t$, and choose $A$ such that the energy of the system reaches the same value at $t=\tau$.  Then using Eq.~\eqref{eq:contin_Fmagdefn} we compare $\bar{F}^{2}$ for both optimal and sinusoidal forcing.  We expect that the optimal forcing is able to cause the system to reach the desired energy at $t=\tau$ using less overall effort.  When we plot the ratio in Fig.~\ref{fig:contin_compare} for a range of $\omega$, we see that it is always less than $1$, as expected.

\begin{figure}[htb]
\centering
 \psfrag{ome}{$\omega$}
 \psfrag{ratio}{$E/\bar{F}^{2}$}
 \psfrag{(a)}{(a)}
 \psfrag{(b)}{(b)}
 \psfrag{k=0}{$k=0$}
 \psfrag{k=1}{$k=1$}
\psfrag{1.8735}{1.8735}
\psfrag{1.8740}{1.8740}
\psfrag{1.8745}{1.8745}
\psfrag{1.8750}{1.8750}
\psfrag{1.8755}{1.8755}
\psfrag{4.97}{4.97}
\psfrag{4.98}{4.98}
\psfrag{4.99}{4.99}
\psfrag{5.00}{5.00}
\psfrag{5.01}{5.01}
\psfrag{1.98}{1.98}
\psfrag{2.00}{2.00}
\psfrag{2.02}{2.02}
\psfrag{2.04}{2.04}
    \includegraphics[width=0.45\textwidth]{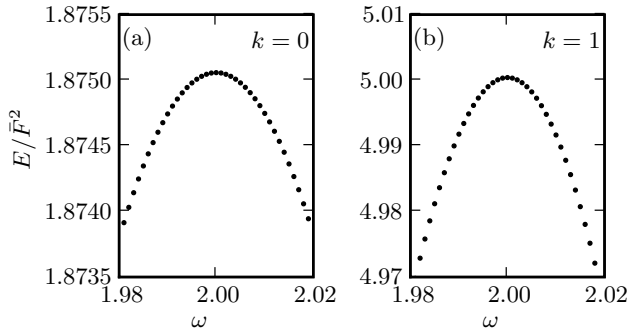}
    \caption{Resonance curve for isotonic harmonic oscillator.  As expected, the ratio of response to total forcing effort is maximal when the natural frequency of the forcing function matches that of the test system ($\omega=\omega_{0}$).  Here, $x(0) = 1.0$, $\dot{x}(0)=0.01$, $\eta = 0.3$, $\omega_{0}=2.00$, and $E-3.6$.  (a) Simple harmonic oscillator with $k=0$.  (b) Isotonic harmonic oscillator with $k=1$.  \label{fig:contin_harmonic_rescurve}}
\end{figure}

At this point we remove the restriction that $k=0$ and consider the system with a nonperturbative nonlinearity ($k\propto \omega$).  Since optimal forcing functions for maps have been demonstrated to be useful for resonance spectroscopy~\cite{gintautas08}, we will show this to be the case for the forcing functions we have calculated above.  For a harmonic oscillator (with $k=0$), the optimal forcing was calculated as a function of the natural frequency $\omega$.  Then a test system harmonic oscillator with $\omega=\omega_{0}$ was forced with this function for different values of $\omega$ until the energy of the oscillator reached the desired value $E$.  The forcing effort $\vec{F}(t)\cdot\vec{F}(t)$ was then integrated from $t=0$ to $t=\tau$ to obtain the total forcing effort $\bar{F}^{2}$.  As expected, the ratio of response to total forcing effort is maximal when the natural frequency of the forcing function matches that of the test system [see Fig.~\ref{fig:contin_harmonic_rescurve}(a)].  We repeat this analysis with an isotonic harmonic oscillator with a nonlinearity, that is, $k\neq0$, and observe similar results [see Fig.~\ref{fig:contin_harmonic_rescurve}(b)].  Thereby the optimal forcing in terms of an unknown and variable parameter may be used for system identification.

\subsection{Explicit example: linear ODE system}
The method presented may also be applied to non-Hamiltonian systems of ordinary differential equations (ODEs).  Since any ODE system can be written as an equivalent first-order system, the method is very general.  We illustrate this with a simple example.  We consider the following system:
\begin{equation}
\begin{pmatrix}
\dot{x}_{1}\\
\dot{x}_{2}
\end{pmatrix}=\begin{pmatrix}
    a x_{1} + k  x_{2}\\
    k x_{1} + a x_{2}
\end{pmatrix}+\begin{pmatrix}
F_{1}\\F_{2}\end{pmatrix}.
\label{eq:contin_linearsystem}
\end{equation}
and require that only $x_{2}$ be forced, that is, $d_u=1$.  The Jacobi matrix of this system is constant and symmetric:
\begin{equation}
    \mathbf{J} = \mathbf{J}^{T}=\begin{pmatrix}
    a& k\\
    k& a 
\end{pmatrix}.
\end{equation}
The method gives rise to the following equations of motion:
\begin{align}
    &\dot{\gamma}=- a \gamma-k F_{2},\\
    &\dot{F}_{2}=- k \gamma - a F_{2}.
\end{align}
Consider a system in which one degree of freedom, $x_{2}$, is accessible to forcing but is coupled to another degree of freedom, $x_{2}$, over which we have no control.  $x_{1}$ may represent, say, the contact age degree of freedom in a sliding friction model~\cite{persson97}.  Suppose we want $x_{2}$ to reach some desired value at $t=\tau$ but have no control over $x_{1}$.  In such a case, at $t=\tau$ we have the simple boundary condition,
\begin{equation}
    K\bigl[\vec{x}(t),\dot{\vec{x}}(t),t\bigr]\equiv x_{2}(t) - C_{p} = 0.
\end{equation}
We will use this boundary condition for this simple example of two coupled linear ordinary differential equations.  Then Eqs.~\eqref{eq:contin_UBeq2_G} and \eqref{eq:contin_UBeq3_G} give rise to the following explicit boundary conditions on $\gamma$ and $F_{2}$, as well as the explicit value of $\lambda$:
\begin{align}
    &\gamma(\tau) = 0,\\
    &F_{2}(\tau) = -2 \bigl[a x_{2}(\tau) + k x_{1}(\tau) \bigr]\\
    &\lambda = -F_{2}(\tau) = 2 \bigl[a C_{p} + k x_{1}(\tau) \bigr]
\end{align}
It is possible to solve the corresponding boundary value problem analytically.  We write the explicit form of $F_{2}(t)$ in terms of the initial conditions,
\begin{align}
    F_{2}(t) &= -2 e^{-at}\bigl({\rm sech}\, kt\bigr)^{2}\cosh{k(t-\tau)} \nn
    &\times \Bigl\{\bigl[k x_{1}(0)+ a x_{2}(0)\bigr] \cosh{kt}\nn
    &+ \bigl[a x_{1}(0) + k x_{2}(0)\bigr]\sinh{kt}\Bigr\},
\end{align}
and we also find an explicit expression for $\bar{F}^{2}$,
\begin{align}
    \bar{F}^{2} &= \frac{({\rm sech}\, kt)^{4}}{2 (a^{3} - ak^{2})} 
    \Bigl\{ \bigl[k x_{1}(0)+ a x_{2}(0)\bigr] \cosh{kt}\nn
    &+ \bigl[a x_{1}(0) + k x_{2}(0)\bigr]\sinh{kt} \Bigr\}^{2}\nn
    &\times \Bigl\{a^{2}-k^{2} - e^{-2at}\bigl(2 a^{2}-k^{2}\bigr)\nn
    &+a^{2}\cosh{2 kt}-ak\sinh{2kt} \Bigr\}.
\end{align}

\begin{figure}[htb]
\centering
 \psfrag{k}{$k$}
 \psfrag{ratio}{$C_{p}/\bar{F}^{2}$}
 \psfrag{0.060}{0.060}
 \psfrag{0.065}{0.065}
 \psfrag{0.070}{0.070}
 \psfrag{0.075}{0.075}
 \psfrag{0.080}{0.080}
 \psfrag{0.0}{0.0}
 \psfrag{0.2}{0.2}
 \psfrag{0.4}{0.4}
 \psfrag{0.6}{0.6}
 \psfrag{0.8}{0.8}
 \psfrag{1.0}{1.0}
 \psfrag{1.2}{1.2}
    \includegraphics[width=0.45\textwidth]{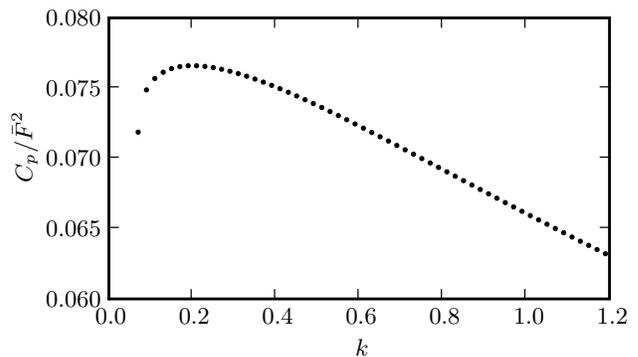}
    \caption{Resonance curve for linear system of first order differential equations.   As expected, the ratio of response to total forcing effort is maximal when the coupling parameter of the forcing function matches that of the test system ($k=k_{0})$.  Here, $x_{1}(0) = 1.0$, $x_{2}(0)=4.0$, $a = 1.3$, $k_{0} = 0.2$, and $C_{p}=1.5$.  \label{fig:contin_linear_rescurve}}
\end{figure}

As with the isotonic harmonic oscillator, we may use the calculated forcing function for resonance spectroscopy.  For this system the optimal forcing was calculated as a function of the coupling parameter $k$.  Then a test system with $k=k_{0}$ was forced with this function for different values of $k$ until $x_{2}$ reached the value $C_{p}$.  The forcing effort $\vec{F}(t)\cdot\vec{F}(t)$ was then integrated from $t=0$ to $t=\tau$ to obtain the total forcing effort $\bar{F}^{2}$.  As expected, the ratio of response to total forcing effort is maximal when the coupling parameter of the forcing function matches that of the test system (see Fig.~\ref{fig:contin_linear_rescurve}).  Just as before, the optimal forcing in terms of an unknown and variable system parameter can be used for system identification.

\section{Conclusions}
We study resonances of forced systems of ordinary differential equations.  We use a constraint at terminal time [Eq.~\eqref{eq:contin_Kcondition}] and seek the forcing function which minimizes the total effort [Eq.~\eqref{eq:contin_Fmagdefn}], subject to the additional constraint that certain degrees of freedom are not directly forced [Eq.~\eqref{eq:contin_fzero}].  To determine this forcing function, we seek the stationary points of the Lagrange function [Eq.~\eqref{eq:contin_Ldefn}] and thereby obtain equations which determine the dynamics of the forcing function [Eqs.~\eqref{eq:contin_GandJ}--\eqref{eq:contin_UBeq3_G}].  From these equations we identify the effective total forcing to be a vector comprising the direct forcing and the Lagrange multipliers that represent the effective indirect forcing of certain degrees of freedom [Eq.~\eqref{eq:contin_Gdefn}].  We demonstrate that the product of the effective forcing and the displacement of nearby trajectories is a conserved quantity [Eq.~\eqref{eq:contin_conserved}].  The methodology presented can be applied to a very general class of problems in which not all of the degrees of freedom in an experimental system are accessible to forcing.  Furthermore, the methodology is not restricted to Hamiltonian systems or systems with small forcing but can applied to any system of ordinary differential equations.

We demonstrate the effectiveness of the methodology with several examples.  We compare forcing calculated using a variational principle to sinusoidal forcing for a damped harmonic oscillator and find that the sinusoidal forcing is less efficient (see Fig.~\ref{fig:contin_compare}).  We present a resonance curve for a damped harmonic oscillator as well as a nonlinear isotonic harmonic oscillator in Fig.~\ref{fig:contin_harmonic_rescurve} and verify explicitly that the optimal effective forcing complements the separation of nearby trajectories [Eq.~\eqref{eq:contin_explicitP}].  We also apply this method to a forced linear system of first order differential equations [Eq.~\eqref{eq:contin_linearsystem}].  We solve for the exact optimal forcing as a function of the terminal time analytically demonstrate that the solution gives the correct peak in the resonance curve (see Fig.~\ref{fig:contin_linear_rescurve}).  Thus we show that the method may be used for system identification.  In the future we plan to compare the effectiveness of this methodology for system identification to that of other methods such as periodic driving~\cite{ruelle86} and coupling a test system to a virtual model with tunable parameters~\cite{gintautas07}.  The method we present need not be restricted to examples such as these.  In fact, the results are general and may be used to implement optimized control of any or all degrees of freedom in systems of ordinary differential equations.

\acknowledgments
The authors wish to thank B. Wah and L. Bettencourt for helpful input and M. Ham, J. Frankel, and A. Gutfriend for helpful discussions.  This work was supported by the National Science Foundation Grant Nos. NSF PHY 01-40179, NSF DMS 03-25939 ITR, and NSF DGE 03-38215.
\appendix*
\section{Variational principle with free terminal time}
Here we derive the equations of motion for a variational problem of the form given in Eq.~\eqref{eq:contin_varproblem}:
\begin{equation}
    \delta S=\delta\int_{0}^{\tau}L(x^{i},\dot{x}^{i},t)+\lambda K(x^{i},\dot{x}^{i},t)\delta_{D}(t-\tau)dt,
\end{equation}
where $x^{i}$ represent generalized coordinates, with $i=1,\ldots,N$.  Since the terminal time $\tau$ is not fixed but is a free parameter, we use a parametric representation of the problem.  Accordingly we replace $t$, $x^{i}$, and $\dot{x}^{i}$ with the following substitution rules:
\begin{align}
    &t=t(p) \qquad\qquad\text{with $t(0)=0$ and $t(1)=\tau$},\\
    &x^{i}(t)=x^{i}\bigl[t(p)\bigr]=x^{i}(p),\\
    &\dot{x}^{i}(t)=\frac{x^{i}_{p}}{t_{p}},
\end{align}
where a subscripted $p$ indicates the partial derivative with respect to the parameter $p$.  By the scaling law for Dirac delta functions,$\int_{0}^{\tau} K\delta_{D}\bigl[t(p)-\tau\bigr]dt = \int_{0}^{1} K\delta_{D}\bigl[t(p)-\tau\bigr]t_{p}dp = \int_{0}^{1} K\delta_{D}(p-1)dp$.  Thus the functional then assumes the form
\begin{equation}
    \delta S=\delta\int_{0}^{1}L(x^{i},\frac{x^{i}_{p}}{t_{p}},t)t_{p}+\lambda K(x^{i},\frac{x^{i}_{p}}{t_{p}},t)\delta_{D}(p-1)dp.
\end{equation}
Then we execute the variation for each variable:
\begin{align}
   &\delta S = \sum_{i=1}^{N}\Biggl(\int_{0}^{1}dp\Biggl[\frac{\partial L}{\partial x^{i}}t_{p}
    + \lambda \frac{\partial K}{\partial x^{i}}\delta_{D}(p-1)\Biggr]\delta x^{i} \nn
   &+ \Biggl[\frac{\partial L}{\partial x_{p}^{i}}t_{p} 
    + \lambda \frac{\partial K}{\partial x_{p}^{i}}\delta_{D}(p-1)\Biggr]\delta x_{p}^{i} \Biggr)\nn
   &+ \Biggl[\frac{\partial L}{\partial t}t_{p} 
    + \lambda \frac{\partial K}{\partial t}\delta_{D}(p-1)\Biggr]\delta t \nn
   &+ \Biggl[\frac{\partial L}{\partial t_{p}}t_{p} 
    + \lambda \frac{\partial K}{\partial t_{p}}\delta_{D}(p-1)\Biggr]\delta t_{p} = 0.
\label{eq:contin_bigvariation}
\end{align}
Next we evaluate the Dirac delta functions and use integration by parts to eliminate $\delta x_{p}^{i}$ and $\delta t_{p}^{i}$.  From Eq.~\eqref{eq:contin_bigvariation} we obtain:
\begin{align}
    \delta S &= \sum_{i=1}^{N}\Biggl\{
    + \Biggl( \lambda \frac{\partial K}{\partial x^{i}}\delta x^{i} \Biggr)\Biggr|_{p=1}
    + \int_{0}^{1}\Biggl(\frac{\partial L}{\partial x^{i}}t_{p}\Biggl)\delta x^{i}dp\nn
   &+ \Biggl( \lambda \frac{\partial K}{\partial x_{p}^{i}}\delta x_{p}^{i} \Biggr)\Biggr|_{p=1}
    + \Biggl[\Biggl(\frac{\partial L}{\partial x_{p}^{i}}t_{p}\Biggl)\delta x^{i}\Biggr]_{0}^{1}\nn
   &- \int_{0}^{1}\frac{d}{dp}\Biggl(\frac{\partial L}{\partial x_{p}^{i}}t_{p}\Biggl)\delta x^{i}dp
    + \Biggl( \lambda \frac{\partial K}{\partial t}\delta t \Biggr)\Biggr|_{p=1}\nn
   &+ \int_{0}^{1}\Biggl(\frac{\partial L}{\partial t}t_{p}\Biggl)\delta tdp
    + \Biggl( \lambda \frac{\partial K}{\partial t_{p}}\delta t_{p} \Biggr)\Biggr|_{p=1}\nn
   &+ \Biggl[\Biggl(L+\frac{\partial L}{\partial t_{p}}t_{p}\Biggl)\delta t\Biggr]_{0}^{1}
    - \int_{0}^{1}\frac{d}{dp}\Biggl(L+\frac{\partial L}{\partial t_{p}}t_{p}\Biggl)\delta tdp = 0.
\end{align}
This gives rise to the following Euler-Lagrange equations for $0<p<1$:
\begin{align}
    &t_{p}\frac{\partial L}{\partial x^{i}} 
    - \frac{d}{dp}\biggl(t_{p}\frac{\partial L}{\partial x_{p}^{i}}\biggr) = 0,\label{eq:contin_pELE1}\\
    &t_{p}\frac{\partial L}{\partial t} 
    - \frac{d}{dp}\biggl(L+t_{p}\frac{\partial L}{\partial t_{p}}\biggr) = 0,\label{eq:contin_pELE2}
\end{align}
for all $i = 1,\ldots,N$.  At the upper boundary, that is, for $p=1$, we have 
\begin{align}
    &\frac{\partial K}{\partial t_{p}} = 0,\label{eq:contin_pUBC1}\\
    \lambda &\frac{\partial K}{\partial t} +L + t_{p}\frac{\partial L}{\partial t_{p}}= 0,\label{eq:contin_pUBC2}\\
    &\frac{\partial K}{\partial x^{i}_{p}} = 0,\label{eq:contin_pUBC3}\\
    \lambda &\frac{\partial K}{\partial x^{i}} + t_{p}\frac{\partial L}{\partial x^{i}_{p}} = 0,\label{eq:contin_pUBC4}
\end{align}
for all $i = 1,\ldots,N$.  At the lower boundary, that is, for $p=0$, we obtain for the variables where the initial conditions $x^{i}(0)$ and $\dot{x}^{i}(0)$ are not fixed:
\begin{align}
    &t_{p}\frac{\partial L}{\partial x^{i}_{p}} = 0.\label{eq:contin_pLBC}
\end{align}
Now we transform back to a parameter-free representation with the following substitutions:
\begin{align}
    &x^{i}(p)=x^{i}\bigl[t(p)\bigr]=x^{i}(t),\\
    &x^{i}_{p}=\dot{x}^{i}\bigl[t(p)\bigr]t_{p} = \dot{x}^{i}(t)t_{p},
\end{align}
for all $i = 1,\ldots,N$.  Eqs.~\eqref{eq:contin_pELE1} and \eqref{eq:contin_pELE1} yield the same equation of motion:
\begin{equation}
    \frac{\partial L}{\partial x^{i}} - \frac{d}{dt} \biggl(\frac{\partial L}{\partial \dot{x}^{i}} \biggr) = 0
    \label{eq:contin_tEOM}
\end{equation}
for all $i = 1,\ldots,N$.   At the upper boundary, that is, for $t=\tau$, we obtain from Eqs.~\eqref{eq:contin_pUBC1}, \eqref{eq:contin_pUBC2}, \eqref{eq:contin_pUBC3}, and \eqref{eq:contin_pUBC1},
\begin{align}
    \lambda &\frac{\partial K}{\partial t} +L - \sum_{i=1}^{N}\biggl[\dot{x}^{i}\frac{\partial L}{\partial \dot{x}^{i}}\biggr]= 0,\label{eq:contin_tUBC1}\\
    &\frac{\partial K}{\partial \dot{x}^{i}} = 0,\label{eq:contin_tUBC2}\\
    \lambda &\frac{\partial K}{\partial x^{i}} + \frac{\partial L}{\partial \dot{x}^{i}} = 0,\label{eq:contin_tUBC3}
\end{align}
for all $i = 1,\ldots,N$.   At the lower boundary, that is, for $t=\tau$, for the variables where the initial conditions $x^{i}(0)$ and $\dot{x}^{i}(0)$ are not fixed, we obtain from Eq.~\eqref{eq:contin_pLBC}:
\begin{equation}
    \frac{\partial L}{\partial \dot{x}^{i}} = 0,\label{eq:contin_tLBC}
\end{equation}
Now we can use the Lagrange function given in Eq.~\eqref{eq:contin_lagrangian} and substitute the variables $x^{i}$ in Eqs.~\eqref{eq:contin_tEOM}--\eqref{eq:contin_tLBC} by $x^{1},\ldots x^{d} = x^{1},\ldots x^{d}$ and $x^{d+1},\ldots x^{2d} = F^{1},\ldots F^{d}$, with $N=2d$.


\end{document}